\documentstyle[twocolumn,aps]{revtex}

\input epsf

\newcommand{\dd}{\mbox{d}}

\newcommand{\ii}{\mbox{i}}
\newcommand{\Li}{{\mbox{Li}_2}}

\newcommand{\gsim}{\mathrel{\raise.3ex\hbox{$>$\kern-.75em\lower1ex\hbox{$\sim$}}}}
\newcommand{\lsim}{\mathrel{\raise.3ex\hbox{$<$\kern-.75em\lower1ex\hbox{$\sim$}}}}

\begin{document}

\draft

\title{
    QED correction to radiative tail from elastic peak in DIS
}

\author{Igor~Akushevich\thanks{on leave
of absence from the National
Center of Particle and High Energy Physics,
220040 Minsk, Belarus}
}
\address{North Carolina Central University,
Durham, NC 27707, USA \\ and \\
TJNAF, Newport News, VA 23606, USA}
\author{Eduard A.~Kuraev, Binur G.~Shaikhatdenov\thanks{on leave
from the Institute of Physics and Technology, Almaty-82 }}
\address{ JINR, 141980 Dubna, Russia }


\maketitle




\begin{abstract}
We calculate DIS cross section under kinematical
requirements when a final hadron state is a single proton.
The process in the lowest order is known as radiative tail from
elastic peak. We take into account correction, coming from emission
of virtual and soft real photons, to the amplitude of this process
as well as the one, induced by emission of additional hard photon
and a hard pair. Resulting contributions of these channels are presented
in the leading and, partly, next-to-leading logarithmic approximations.
Some general expressions for the photon contributions are given.
Numerical results are presented for kinematical conditions of the
current experiments on DIS.
\end{abstract}

\narrowtext

\section{Introduction}
\label{sec:Intro}

Deep inelastic scattering (DIS) is one of the powerful tools in
investigating a nucleon nature. Recent experiments at CERN, DESY
and SLAC have provided very precise data in the wide kinematical
region. The modern level of data analysis in experiments on DIS
requires careful consideration of the QED radiative effects which can give
substantial contribution to measured quantities. Usually
the radiative photon cannot be registered in a
detector. As is well understood the corrections due to soft photons
and loop effects cannot be
separated from observables  in principle. Hence their
contribution has to be calculated theoretically and subtracted from
observed data. The lowest order radiative corrections were first
calculated by Mo and Tsai~\cite{MoTsai}. Covariant approach utilized
in~\cite{BSh} was applied to unpolarized~\cite{ABSh1} and
polarized~\cite{KSh,ASh} DIS.
Second order correction to DIS cross section in the
leading approximation  is discussed in~\cite{FKM,heavy,AAK} and recently
in~\cite{AISh}. For completeness we cite the papers~\cite{Spies,Bardin}
in which the correction was calculated within a framework of
electroweak theory basically for HERA
kinematics. The excellent review of the lowest order RC can be found, say,
in~\cite{ABKR}.

One of the important contribution to RC comes from a so called radiative
tail from elastic peak (or simply elastic radiative tail),
 when final hadronic state is pure nucleon but the invariant
mass of unobserved
particle system (a radiated photon and this nucleon) is the same
as in the main Born-level process. Consequently events detected
can correspond to
the main process as well as to the elastic radiative tail. The
calculation in the lowest order of QED is standard and the results are
included in many FORTRAN codes intending to perform the RC procedure
of experimental data (see review~\cite{HERA} and more recently
developed codes~\cite{HERACLES,HECTOR,POLRAD,RADGEN}). Several papers
\cite{AK,ABK} were devoted to electroweak correction to elastic
radiative tail.
 Numerical analysis of the elastic radiative tail
shows that its contribution is very important and can exceed the main
measured process at the Born level. Therefore the next step is to calculate
QED correction to the elastic radiative tail  with the maximal
possible accuracy. So far only	the leading
 correction to elastic radiative tail due to double bremsstrahlung, which
is part of the total second order correction, was treated numerically
\cite{AISh,POLRAD} and the attempt to calculate the correction exactly
was done in~\cite{ABSh2}.

The structure of the cross section of elastic radiative
tail is the following
\begin{equation}
\sigma^{ERT} \sim \int\limits_{Q^2_{h\;min}} \dd Q^2_{h} {\cal
  K}(Q^2_h,Q^2,W^2)
{\cal F}^2(Q^2_h),
\end{equation}
where $Q^2_h$ is a momentum square transferred to hadronic
system, and $Q^2$ and $W^2$
are leptonic kinematical variables measured. The quantity $\cal K$
is a kinematical
factor known exactly and $\cal F$ is a nuclear form factor. Due to rapid
fall of the form factor squared as a function of $Q^2_h$ the main
integration region is close to the lower integration limit.
In papers~\cite{AK,AISh} this fact was used to construct an
approximation where $Q^2_h$ is considered as a small parameter of
order of the proton mass squared. In this paper we will also use
this approximation to analyze the correction to elastic radiative
tail. Application of Sudakov technique will allow us to obtain compact
explicit formulae for processes considered.
The first effect which has  to be considered is the one-loop correction
and the emission of additional real photon. We will analyze both of them at
leading and next-to-leading levels. Another effect contributing to
the cross section is a lepton pair creation.
We will calculate it in the leading log approximation.

Obtaining a second order correction to deep inelastic scattering is the main
motivation of this paper. However our results can be used in other
cases. For instance, they can be considered as a radiative
correction in measurements with hard photon detected
in coincidence with scattered
lepton (see~\cite{coinc}, for example), that allows one to reach
kinematical regions otherwise unreachable. That is why we do not
concretize our notation usually used in DIS but instead try to keep
it as general as possible. In the next section we introduce our notation
and obtain results for the cross section of single bremsstrahlung
using Sudakov technique. In section~\ref{sec:VS} we give results for
one-loop corrections. Double bremsstrahlung and contributions due to
pair production are considered in sections~\ref{sec:2h}
and \ref{sec:pair} and final remarks are given in
section~\ref{sec:con}. Some technical details are discussed
in Appendices.

\section{Single bremsstrahlung}
\label{sec:SB}


We study the process
\begin{eqnarray}\label{eq:1}
&&e(p_1)+p(p_2)\to e(p_1')+\gamma(k_1)+p(p_2'),\quad s=2p_1p_2,\nonumber \\
&&Q_h^2=-(p_2-p_2')^2, \quad Q^2=2p_1p_1', \quad k_1^2=0, \\ \nonumber
&&p_1^2=p_1^{{'}2}=m^2,\quad p_2^2=p_2^{2'}=M^2,\quad q^2=-O^2_h\,,
\end{eqnarray}
in the kinematical region
\begin{equation}
s\gg Q^2>Q_h^2\sim M^2,\qquad 2p_2p_1'\sim s.
\end{equation}
The expression for differential cross section in Born approximation looks
(details are given in the Appendix A):
\begin{eqnarray}\label{eq:4}
2\varepsilon_1'\frac{\dd^3\sigma_0^\gamma}{\dd^3p_1'}
=\frac{4\alpha^3}{\pi^2}\int\frac{\dd^2\mathbf{q}}
{({\mathbf q}^2+Q^2_{\mathrm{min}})^2}\frac{1}{1-b}
\Phi^\gamma \Phi^{\mathrm{prot}},
\end{eqnarray}
with $b=2p_2p_1'/s$ the energy fraction of the scattered electron.
We imply the Sudakov parameterization of the 4-momenta in the problem
(see Appendix A).

Note that due to gauge invariance condition
\begin{equation}
q^\rho J^{(1)}_\rho\approx(\alpha_q p_2+q_\bot)^\rho J^{(1)}_\rho=0,
\end{equation}
the quantity $\Phi^\gamma$ is constructed out of $(1/s)p_2J^{(1)}$
which may be rearranged as follows:
\begin{eqnarray}
\frac{1}{s}p_2^\mu J^{(1)}_\mu&=&-\frac{s}{s_1}|{\mathbf{q}}|e_q^\mu J^{(1)}_\mu,
\quad e_q=\frac{\mathbf{q}}{|\mathbf{q}|}, \\ \nonumber
s_1&=&s\alpha_q=(p_1'+k_1)^2+{\mathbf{q}}^2-m^2.
\end{eqnarray}
Thus $\Phi^\gamma$ vanishes for small ${\mathbf{q}}^2$. The explicit
expression for $\Phi^{\mathrm{prot}}$ is found to be
\begin{equation}
\Phi^{\mathrm{prot}} = 2(F_1^2 + \frac{{\mathbf{q}}^2}{M^2}F_2^2).
\end{equation}
For $\Phi^\gamma$ we have (we refer for further details to the Appendix A):
\begin{eqnarray}\label{eq:8}
\Phi^\gamma = \frac{(1-b)^2 b(1+b^2){\mathbf{q}}^2}{n_1n}\,, \\ \nonumber
\end{eqnarray}
with
\begin{eqnarray}
n = ({\mathbf{p}}_1' - b{\mathbf{q}})^2,\quad
n_1 = ({\mathbf{p}}_1' - {\mathbf{q}})^2. 
\end{eqnarray}
Another fact is that both $\Phi^\gamma/{\mathbf{q}}^2$
and $\Phi^{\mathrm{prot}}$
do not vanish in the limit of small momentum transfer $|{\mathbf{q}}|$,
thus providing the
logarithmic enhancement upon performing the $Q^2_h\approx {\mathbf{q}}^2$
integration (Weizs\"acker-Williams approximation). Indeed,
the quantity $Q^2_{\mathrm{min}}$ entering the cross section is
a small quantity,
\begin{equation}
Q^2_{\mathrm{min}}=M^2\left(\frac{Q^2}{(1-b)s}\right)^2\ll M^2.
\end{equation}
For completeness we put the phase volume of the scattered electron
in terms of Sudakov variables:
\begin{eqnarray}
\frac{\dd^3p_1'}{2\varepsilon_1'} = \frac{\dd b}{2 b}
\dd^2{\mathbf{p}}_1', \quad Q^2 = 2p_1p_1' = \frac{{\mathbf{p}}_1^{'2}}{b}.
\end{eqnarray}
Note that the requirement $Q^2>Q_h^2$ provides the absence of singularities
while doing an integration over $\dd^2{\mathbf{q}}$.


\section{Virtual and soft photons emission contribution}
\label{sec:VS}

The correction coming from the emission of virtual and soft photons
(in the cms reference frame) can be drawn out of
paper~\cite{heavy}, in which the radiative
corrections to the Compton tensor were calculated
\begin{eqnarray} \label{loops}
2\varepsilon_1'\frac{\dd^3\sigma_{B+V+S}}{\dd^3p_1'}
\!\!&=&\!\!2\varepsilon_1'\frac{\dd^3\sigma_0^\gamma}
{\dd^3p_1'}\Biggl[1+\frac{\alpha}{2\pi}\tilde\rho \\ \nonumber
\!\!&+&\!\! \frac{\alpha}{4\pi}\frac{1}{1+b^2}\biggl(\tau_{11}+b(\tau_{12}
+\tilde{\tau}_{12}) + b^2\tilde{\tau}_{11}\biggr)\Biggr],
\end{eqnarray}
with
\begin{eqnarray}
\tilde\rho&=&2(L-1)(2\ln\Delta-\ln b)+3L_h-\ln^2b \\ \nonumber
&-&\frac{9}{2}
-\frac{\pi^2}{3} + 2\Li\left(\cos^2\frac{\theta}{2}\right), \\ \nonumber
L&=&\ln\frac{Q^2}{m^2},\qquad L_h=\ln\frac{Q_h^2}{m^2}, \\ \nonumber
\Delta&=&\frac{\Delta E}{E},\qquad
\Li(x)=-\int\limits_{0}^{x}\frac{\ln(1-y)}{y}\dd y.
\end{eqnarray}
The Born cross section after substitution of Eq.~(\ref{eq:8}) into
Eq.~(\ref{eq:4}) and neglect of sub-leading terms becomes
\begin{eqnarray*}
2\varepsilon_1'\frac{\dd^3\sigma_0^\gamma}{\dd^3p_1'}
=\frac{4\alpha^3}{\pi^2}\int\frac{\dd^2 {\mathbf{q}}{\mathbf{q}}^2}
{({\mathbf{q}}^2
+ Q_{\mathrm{min}}^2)^2}\frac{(1-b)(1+b^2)}{b(Q^2)^2}\Phi^{\mathrm{prot}},
\end{eqnarray*}
where $\Delta E, E$ are the upper bound on the undetectable soft photon
energy, and the energy of the initial electron, respectively;
$\theta$ is the angle
in the laboratory reference frame between the initial and the scattered
electron momenta. Somewhat cumbersome functions $\tau_{ij}$
are explicitly given in the Appendix D. It should
be noted that they do not contain any large logarithms but include the
quantity $Q^2_h$ which is small in our approximation. If one
keeps only non-zero terms in the expansion over $Q^2_h$ then
\begin{eqnarray}\label{tata}
&& \frac{1}{2}\biggl(\tau_{11}+b(\tau_{12}
+\tilde{\tau}_{12}) + b^2\tilde{\tau}_{11}\biggr)
=\biggl[3\log {Q^2\over Q^2_h(1-b)} - 1\biggr] \nonumber\\
&&\qquad \times(1+b^2)
+ 4b\log (1-b) + [b^2 + (1-b)^2] \nonumber \\
&&\qquad \times\biggl[\log^2{(1-b)\over b}+\pi^2\biggr]
+ [1+(1-b)^2]\log^2 (1-b) \nonumber \\
&&\qquad +(3-2b)\log b,
\end{eqnarray}
The logarithms $\log Q^2_h$  cancel out
exactly in the sum of (\ref{tata}) and $\tilde\rho$.

\section{Two hard photons emission contribution}
\label{sec:2h}

We will consider now the process of two hard
photons emission:
\begin{equation}
e(p_1)+p(p_2)\to e(p_1')+\gamma(k_1)+\gamma(k_2)+p(p_2').
\end{equation}
The relevant contribution to the cross section looks
\begin{eqnarray}
2\varepsilon_1'\frac{\dd^3\sigma}{\dd^3p_1'}
&=&\frac{\alpha^4}{8\pi^4}\int\frac{\dd^2{\mathbf{q}}}
{({\mathbf{q}}^2+Q^2_{\mathrm{min}})^2}\frac{\dd x_1 \dd^2{\mathbf{k}}_1}
{x_1x_2}\Phi^{\gamma\gamma} \Phi^{\mathrm{prot}},
\nonumber  \\
Q^2_{{\mathrm{min}}}&=&M^2\left(\frac{s_1}{s}\right)^2,\quad
s_1=(p_1'+k_1+k_2)^2\,,
\end{eqnarray}
with the expression for $\Phi^{\mathrm{prot}}$ given earlier.
The explicit form of $\Phi^{\gamma\gamma}$ can be found in the Appendix B.
The integration over $\dd^2{\mathbf{k}}_1$ may be performed using the integrals
given in the Appendix C.

Concerning the region $Q_h^2\ll Q^2$, the result is found to be
\footnote{Upon applying the crossing transformation to the amplitude
of $e\bar{e}$ annihilation to $\gamma\gamma\gamma$
presented in paper~\cite{ber}.}
\begin{eqnarray}\label{eq:16}
\Phi^{\gamma\gamma}&=&16{\mathbf{q}}^2\Biggl\{\frac{Q^2}{s_1^2}
\biggl[\frac{s_1^2(1+b^2)}{d_1d_2d_1'd_2'} + \frac{d_1^2 + d_1^{'2}}
{b s_1^2 d_2 d_2'} + \frac{d_2^2+d_2^{'2}}{b s_1^2 d_1 d_1'}\biggr] \nonumber \\
&-& \frac{2}{Q^4}(1 + {\cal P}_{12})
\biggl[\frac{m^2}{d_1^2}\frac{x_2^2(b^2+(1-x_1)^2)}{b(1-x_1)^3} \\ \nonumber
&+& \frac{m^2}{d_1^{'2}}
\frac{x_2^2 b^2(1+(1-x_2)^2)}{(1-x_2)^3}\biggr]\Biggr\}.
\end{eqnarray}
with the notations introduced
\begin{eqnarray}
&&s_1=\frac{{\mathbf{k}}_1^2}{x_1}+\frac{{\mathbf{k}}_2^2}{x_2}
+ \frac{{\mathbf{p}}_1^{'2}}{b},\qquad
d_i=\frac{1}{x_i}(m^2x_i^2+{\mathbf{k}}_i^2),	\\ \nonumber
&&
d_i'=\frac{1}{x_i b}[m^2x_i^2 + (x_i{\mathbf{p}}_1'-b{\mathbf{k}}_i)^2],
\end{eqnarray}
where $x_{1,2}$ are the energy fractions of hard photons, $x_1+x_2+b=1$.
Besides we use the relations
$$
{\mathbf{k}}_1+{\mathbf{k}}_2+{\mathbf{p}}_1'=0\,,\quad
2qp_1'=s_1 b\,,\quad s_1=2qp_1=s\alpha_q.
$$
An integration over $\dd^2{\mathbf{k}}_1$
may be performed analytically and to a logarithmic accuracy
it boils down to
\begin{equation}
\int\frac{\dd^2{\mathbf{k}}_1}{\pi}
\left[\frac{1}{d_1}; \frac{1}{d_2}; \frac{1}{d_1'};
    \frac{1}{d_2'}\right]
    = L\left[x_1; x_2; \frac{x_1}{b}; \frac{x_2}{b}\right].
\end{equation}
The resulting contribution (again to a logarithmic accuracy) takes
the following form
\begin{eqnarray}
\int \dd^2 {\mathbf{k}}_1 \Phi^{\gamma\gamma}&=&\frac{16\pi{\mathbf{q}}^2 L}{b(Q^2)^2}
(1 + {\cal P}_{12})x_2^2 \nonumber \\
&\times&\Biggl[\biggl(1+\frac{1}{(1-x_1)^2} + \frac{b^2}{(1-x_2)^2}\biggr)
(1+b^2)  \nonumber \\
&+& \frac{b^2}{(1-x_1)^4} + \frac{b^4}{(1-x_2)^4}\Biggr], \\ \nonumber
\Delta&<& x_i < 1 - b - \Delta.
\end{eqnarray}
Carrying out the integration of Eq.~(\ref{eq:16}) over $\vec k_1$ and $x_1$
to a next-to-leading accuracy we obtain for the contribution to
the differential cross section coming from emission of two hard
photons,
\begin{equation}\label{twopho}
2\varepsilon_1'\frac{\dd^3\sigma^{\gamma\gamma}}{\dd^3p_1'}=\frac{2\alpha^4}
{\pi^3}\int
\frac{\dd^2{\mathbf{q}}}{({\mathbf{q}}^2 + Q^2_{\mathrm{min}})^2}
\frac{{\mathbf{q}}^2 (T_{LL}+T_{NLO})}{b(Q^2)^2}\Phi^{\mathrm{prot}}
\end{equation}
where the leading and next-to-leading contributions read
\begin{eqnarray}\label{eq:22}
T_{LL}
&=&(L-1)[4(1-b)(1+b^2)\ln\frac{1-b}{\Delta}
 \nonumber \\
&+&(1-b)(1-b^2)\ln b - \frac{2}{3}(1-b)(7-2b+7b^2)],
 \nonumber \\
T_{NLO}
&=&-\frac{1}{2}{b^4+6b^2+1 \over 1+b}\log^2b
 \nonumber  \\
&&-\frac{1}{3}(3-b^2)(3-b) \log b
\\ \nonumber
&&+\frac{8}{3}(1-b)(b^2+b+1) \log (1-b)
\\ \nonumber
&&-(1-b)\biggl[\frac{1}{3}(15b^2-2b+15)
\\ \nonumber
&&+2\biggl({\rm \Li}(b)-\frac{\pi^2}{6}\biggr){b^4+6b^2+1 \over
1-b^2}\biggr].
\end{eqnarray}

There are two possible experimental setups we concern with: the first one
in which a recoil proton is registered, and the second ---
pure inclusive setup --- with only a final lepton observed.
Definitely, NLO contribution obtained can be counted valid
only for the former experimental setup, while in the latter case
one can use the expression given above only to a LL accuracy.

The general answer for the cross section in Born approximation with the
lowest order correction to the leading approximation is a sum of the
contributions coming from virtual and real soft photons emission given above
as well as from two hard photons emission and is free from dependence
on the auxiliary parameter $\Delta$.

The graphs given below illustrate behavior featured by the complete
QED RC contribution to the cross section of DIS as well as
the comparative contributions of the
LL, NLO terms and of the correction due to pair production.

\section{Contribution of lepton pair production}
\label{sec:pair}

Consider now the hard pair production process that takes place at
the same order of perturbation theory as the two hard photon emission.
In the same
way we may conclude that the soft pair case as well as the case of
double collinear kinematics does not contribute to the radiative tail.
Therefore we may consider only semi-collinear kinematics of hard pair
production of which there exist two different mechanisms~\cite{akmt}.
One of these is the two photon mechanism of pair creation.
An electron from that pair having momentum $p_1'$ is detected
in experiment and the scattered electron moves close to the initial
electron direction.
This kinematics permits us to apply the Weizs\"acker-Williams approximation,
\begin{eqnarray}\label{pair1}
2\varepsilon_1'\frac{\dd^3\sigma^{(1)}_{{\mathrm{pair}}}}
{\dd^3p_1'}&=&\frac{2\alpha^4}{\pi^3}\int
\frac{\dd^2{\mathbf{q}}}{({\mathbf{q}}^2 + Q^2_{\mathrm{min}})^2}
\frac{{\mathbf{q}}^2 L}{b (Q^2)^2}\Phi^{\mathrm{prot}} \\ \nonumber
&\times&\frac{\dd\beta_-}{(1-\beta_-)^4}((1-\beta_- - b)^2+b^2)
(1+\beta_-^2),	\\ \nonumber
s_1&=&Q^2\frac{1-\beta_-}{\beta_+},\qquad b+\beta_-+\beta_+=1.
\end{eqnarray}
The second mechanism is characterized by the bremsstrahlung mechanism
of pair creation, with an electron from a pair to be detected.
Leaving details to the Appendix~E let us present here the result
\begin{eqnarray}\label{pair2}
2\varepsilon_1'\frac{\dd^3\sigma^{(2)}_{{\mathrm{pair}}}}{\dd^3p_1'}
&=&\frac{2\alpha^4}{\pi^3}\int
\frac{\dd^2{\mathbf q}}{({\mathbf q}^2
+ Q^2_{\mathrm{min}})^2}
\frac{{\mathbf q}^2 L}{(Q^2)^2}\Phi^{\mathrm{prot}}  \nonumber \\
&\times&\frac{b(1 + \beta_-^2)\dd\beta_-}{(1-\beta_-)^4}[(1- b
- \beta_-)^2 + b^2] \\ \nonumber
s_1 &=& Q^2\frac{1-\beta_-}{b\beta_-}.
\end{eqnarray}

The integration over $\beta_-$ can be performed analytically with
additional assumption that $Q^2_{\mathrm{min}}$ has no $\beta_-$ dependence.
The result for the sum of these contributions is found to be
\begin{eqnarray}\label{pairsum}
2\varepsilon_1'\frac{\dd^3\sigma_{{\mathrm{pair}}}}{\dd^3p_1'}
&=&\frac{2\alpha^4}{\pi^3}\int
\frac{\dd^2{\mathbf q}}{({\mathbf q}^2
+ Q^2_{\mathrm{min}})^2}
\frac{{\mathbf q}^2 L(1+b^2)}{b(Q^2)^2}\Phi^{\mathrm{prot}}  \nonumber \\
&\times& \biggl( 1-b+2(1+b)\log b+\frac{4}{3b}(1-b^3) \biggr).
\end{eqnarray}

\section{Discussion and conclusion}
\label{sec:con}

In the paper presented the correction to
radiative tail from elastic peak is studied in the kinematics when
a final lepton is measured. Using Sudakov technique
the contributions of loops~(\ref{loops}), double photon
bremsstrahlung~(\ref{twopho})
and a pair production~(\ref{pair1},\ref{pair2}) are calculated.

In this section we analyze obtained contributions numerically.
Both the relative contributions of the processes considered and the total
correction to the lowest order process are investigated
within kinematical conditions of
experiments on electron DIS at TJNAF and DESY (both for HERA and
for HERMES). It is convenient to define the following quantities:
\begin{equation}
\delta={\sigma_L+\sigma_N+\sigma_p \over \sigma_0},\quad
\delta_{L,N,p}={\sigma_{L,N,p}\over \sigma_0}.
\end{equation}
Here $\sigma_0$ stands for the cross section of radiative tail from
elastic peak~(\ref{eq:4}).
Other $\sigma$'s constitute the next order results.
The quantity $\sigma_p$ is a direct sum of two mechanisms of
pair creations (\ref{pair1},\ref{pair2}), whereas $\sigma_L$ and $\sigma_N$
are the leading (including mass singularities terms $\log(Q^2/m^2)$) and
next-to-leading (independent of leptonic mass) terms.

\begin{figure}[t]
\unitlength 1mm
\begin{picture}(80,80)
\put(0,5){
\epsfxsize=8cm
\epsfysize=8cm
\epsfbox{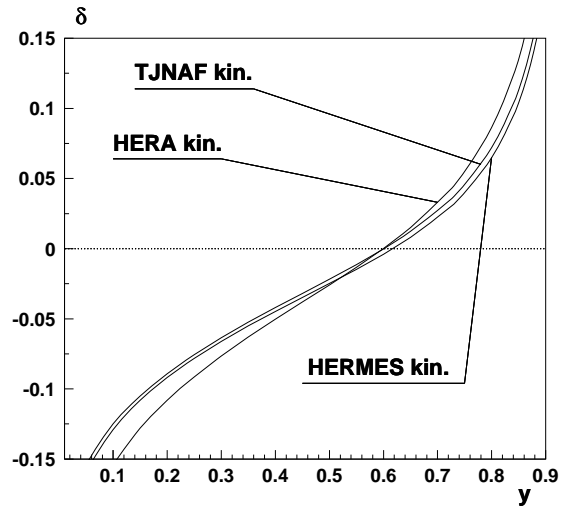}
}
\end{picture}
\caption{QED correction to radiative tail from elastic peak
in electron DIS for the kinematics of
TJNAF ($\sqrt{S}$=4. GeV, $x=$0.05),
HERMES ($\sqrt{S}$=7 GeV, $x=$0.5) and
HERA ($\sqrt{S}$=300 GeV, $x=$0.0005).
}
\label{Fig1}
\end{figure}

\begin{figure}[t]
\unitlength 1mm
\begin{picture}(80,80)
\put(0,5){
\epsfxsize=8cm
\epsfysize=8cm
\epsfbox{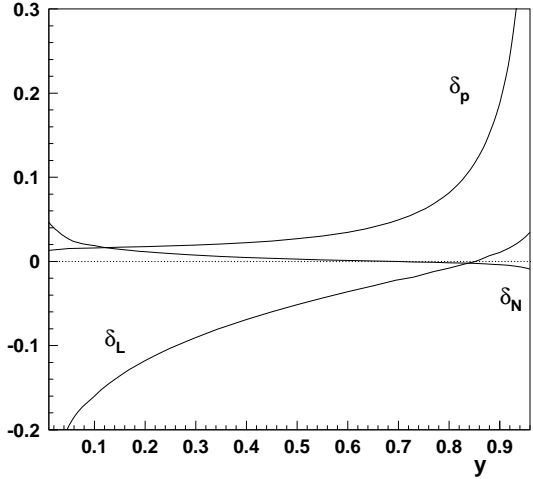}
}
\end{picture}
\caption{The leading ($\delta_L$), next-to-leading
($\delta_N$) contributions of double photon bremsstrahlung and correction
due to pair creation ($\delta_p$).
}
\label{Fig2}
\end{figure}

They are obtained
upon summing up expressions given in Eqs.~(\ref{loops}) and~(\ref{twopho})
after cancellation of infrared divergence.

The quantities $\delta$ and $\delta_{L,N,p}$ are presented in
Figs.~\ref{Fig1} and~\ref{Fig2}. As is clearly seen the corrections
have a steep dependence on $y$
and a poor one on $x$ (or $Q^2$) and $s$. Figure~\ref{Fig2} shows
results for $\sqrt{s}=7\ GeV$, however very similar plots can be produced for
the two other cases considered.
Having analyzed results obtained the following conclusions could be drawn.

The relative radiative correction to elastic radiative tail is
important practically everywhere. The modern level of data analysis and
very high experimental accuracies achieved in current experiments on DIS
require that a generalization of standard radiative correction procedure
be made in order to include a second order radiative correction.
An extremely interesting
region where the correction considered is important is actually
high $y$ domain.
Remind, that this one (up to $y\sim$0.95) is under investigation
at TJNAF.

The main contribution to a second order radiative correction comes from
the effect of pair creation. Asymptotical behavior of $\sigma_p$ for
small $b=1-y$ is $1/b^2$ whereas the other cross sections feature only $1/b$
behavior. That is in fact a reason of the large correction in the region
of high $y$. In the paper presented this particular contribution
is calculated in the leading approximation only,
therefore a study of the correction, induced by a pair production,
at the next-to-leading level is highly desirable.

The relative contribution of the next-to-leading correction $\sigma_N$ is not
small with respect to the leading log contribution $\sigma_L$.
In the region of large $y$ the relative contribution
$\sigma_N/\sigma_L$ does not exceed 5\%, whereas for small $y$
it can reach as much as 20--30\%. From the other
hand the next-to-leading contribution completely fixes all
uncertainties of leading log approximation thus leaving unknown only terms
proportional to lepton mass squared and  $Q^2_h$, which is
effectively small due to behavior of form factors.

\section*{Acknowledgements}

We would like to thank P.Kuzhir for useful discussions and comments.
The work of IA was partially supported by the
U.S. Department of Energy under contract DE--AC05--84ER40150.
EAK and BGS acknowledge support of RFBR via grant No.~99-02-17730.

\section*{Appendix A. Details of matrix element calculus:
the case of single photon bremsstrahlung}
\label{sec:AppA}
\setcounter{equation}{0}
\renewcommand{\theequation}{A.\arabic{equation}}

Using the Sudakov decomposition of the 4-vectors in the problem
\begin{eqnarray}
&&p_1'=\alpha_1'\tilde{p}_2+b\tilde{p}_1+p_{1\bot}',\quad
k_1=\alpha_1\tilde{p}_2+x_1\tilde{p}_1+k_{1\bot},  \nonumber \\
&&q=p_2-p_2'=\alpha_q\tilde{p}_2+\beta_q\tilde{p}_1+q_\bot,
 \nonumber \\
&&p_2'=\alpha_2'\tilde{p}_2+\beta_2'\tilde{p}_1+p_{2\bot}',\quad
v_\bot p_1=v_\bot p_2=0, \\ \nonumber
&&\tilde{p}_1=p_1-p_2\frac{m^2}{s},\quad \tilde{p}_2=p_2 - p_1\frac{M^2}{s},
\end{eqnarray}
we have excluded parameters $\alpha_1,\alpha_1',\beta_q$ using
the on--shell conditions
\begin{eqnarray}
&&p_2^{{'}2}-M^2=-s\beta_q(1-\alpha_q)-{\mathbf{q}}^2-\alpha_q M^2=0,\\ \nonumber
&&p_1^{{'}2}=s b\alpha_1'-{\mathbf{p}}_1^{{'}2}=0,
    \quad k_1^2=s x_1\alpha_1-{\mathbf{k}}_1^2=0,
\end{eqnarray}
besides
\begin{eqnarray}
\Phi^{\mathrm{prot}}&=&\frac{1}{s^2}{\mathrm{Sp}}
\{(\hat{p}_2'+M)\Gamma_\rho(\hat{p}_2+M)
\tilde{\Gamma}_\sigma p_1^\rho p_1^\sigma\}, \nonumber \\
\Gamma_\rho&=&F_1(q^2)\gamma_\rho + \frac{\sigma_{\mu\rho}q^\mu}{2M}F_2(q^2).
\end{eqnarray}
Here $F_{1,2}(q^2)$ are the Dirac and Pauli form factors of a proton.
For $\Phi^\gamma$ we have:
\begin{eqnarray}
&&\Phi^\gamma=-\frac{1}{s^2}{\mathrm{Sp}}\{\hat{p}_1'O_\mu\hat{p}_1
\tilde{O}^\mu\},\nonumber \\
&&O_\mu=\hat{p}_2\frac{\hat{p}_1-\hat{k}_1}{-2p_1k_1}\gamma_\mu+
\gamma_\mu\frac{\hat{p}_1'+\hat{k}_1}{2p_1'k_1}\hat{p}_2,
\end{eqnarray}
and then
\begin{equation}
q^2=-Q_h^2=-\frac{1}{1-\alpha_q}[{\mathbf{q}}^2+M^2\alpha_q^2]\approx -[{\mathbf{q}}^2
+ Q_{\mathrm{min}}^2],
\end{equation}
with $Q_{\mathrm{min}}^2$ given in the text.
The matrix element
\begin{equation}
M=\frac{1}{q^2}J^{(1)}_\sigma \bar{u}(p_2')\Gamma_\rho u(p_2)g^{\rho\sigma},
\end{equation}
using the Gribov representation for the metric tensor
\begin{equation}
g^{\rho\sigma}=g_\bot^{\rho\sigma}+
\left(\frac{2}{s}\right)(\tilde{p}_2^\rho\tilde{p}_1^\sigma+
\tilde{p}_2^\sigma\tilde{p}_1^\rho)
\approx\left(\frac{2}{s}\right)\tilde{p}_2^\sigma\tilde{p}_1^\rho,
\end{equation}
may be put in a form
\begin{equation}  \label{A8}
M=\frac{2s}{q^2}\left(\frac{1}{s}p_2^\sigma J^{(1)}_\sigma\right)
\left(\frac{1}{s}\bar{u}(p_2')\Gamma_\rho u(p_2)p_1^\rho\right).
\end{equation}
Note that each expressions in the parentheses on the RHS of Eq.~(\ref{A8})
do not depend on $s$ in the limit $s\to\infty$.
The expression for $\Phi^\gamma$ may be transformed
using the following reduced expression
\begin{eqnarray}
O_\mu=x_1\biggl[sb\gamma_\mu\left(\frac{1}{n}-\frac{1}{n_1}\right)
&+&\frac{1}{n_1}b\gamma_\mu\hat q\hat p_2
-\frac{1}{n}\gamma_\mu\hat p_2\hat q\biggr], \\ \nonumber
x_1&=&1-b.
\end{eqnarray}
to take the form given in Eq.~(\ref{eq:8}).

\section*{ Appendix B. Details of matrix element calculus:
the case of double photon bremsstrahlung}

\label{sec:AppB}
\setcounter{equation}{0}
\renewcommand{\theequation}{B.\arabic{equation}}

Let's first demonstrate that the matrix element of the process
 \begin{equation}
 \gamma^*(q)+e(p_1)\to e(p_1')+\gamma(k_1)+\gamma(k_2)
 \end{equation}
is explicitly
 proportional to ${\mathbf{q}}$ for small values of the latter,
 which is in fact the requirement of gauge invariance with respect to
 the virtual photon.
 The matrix element is described by six diagrams. With regard to the gauge
 invariance this set can be separated out to the two subsets
 in each of which the gauge condition is
 satisfied independently. Introducing the photon-permutating operator
 ${\cal P}_{12}$ we bring the matrix element to the form:
\begin{eqnarray} \label{eq:12}
  {\cal{M}}=(1+{\cal{P}}_{12})Q \,, \qquad
  Q={\cal{M}}_1+{\cal{M}}_2+{\cal{M}}_3\,,
\end{eqnarray}
where
\begin{eqnarray} \label{amplitudes}
  {\cal{M}}_1&=&\frac{1}{d d_1}\bar {u}(p_1')\hat p_2(\hat p_1-\hat k_1
  -\hat k_2+m) \\ \nonumber
  &&\times\hat e_2^*(\hat p_1-\hat k_1+m) \hat e_1^*u(p_1)\,,
\nonumber
\\
  {\cal{M}}_2&=&\frac{1}{d_1 d_2'}\bar u(p_1')
  \hat e_2^*(\hat p_1-\hat k_1+ \hat q+m)\hat p_2
  \\ \nonumber
  &\times&(\hat p_1-\hat k_1+m)\hat e_1^* u(p_1)\,,
\\
  {\cal{M}}_3&=&\frac{1}{d'd_2'}\bar u(p_1')
  \hat e_2^*(\hat p_1-\hat k_1+\hat q+m) \\ \nonumber
  &&\times\hat e_1^*(\hat p_1+\hat q+m) \hat p_2 u(p_1) \,,
\nonumber
\end{eqnarray}
and
\begin{eqnarray*}
d&=&d_1+d_2-\frac{1}{x_1x_2}(x_1\vec{k}_2 - x_2\vec{k}_1)^2,\\
d'&=&d_1'+d_2'+\frac{1}{x_1x_2}(x_1\vec{k}_2 - x_2\vec{k}_1)^2.
\end{eqnarray*}
The permutation operator ${\cal{P}}_{12}$ for the photons acts
the following way
\begin{eqnarray*}
{\cal{P}}_{12}f(k_1,e_1;k_2,e_2)=
f(k_2,e_2;k_1,e_1)\,, \qquad {\cal{P}}_{12}^2 =1. \nonumber
\end{eqnarray*}

The quantity $Q$ is gauge invariant regarding the virtual photon $k$
since all permutations of this photon have been taken into account.
Therefore $Q$ is proportional to $q_\perp$ in the limit of
$q_\perp \to 0$. Indeed, making use of the relations
\begin{eqnarray}
  Q = p_{2\mu} Q^\mu \,, \ \ \ q_\mu Q^\mu=
  (\alpha_q \tilde p_2 + q_\perp)_\mu Q^\mu =0\,,
\end{eqnarray}
we immediately obtain
(neglecting the small contribution $\beta_q p_\mu Q^\mu\sim 1/s$)
\begin{eqnarray} \label{eq:13}
 Q= -\frac{q_{\perp\mu}}{\alpha_q} Q^\mu \,.
\end{eqnarray}

Then transform the quantities ${\cal M}_j$ to such a form that
the noticed low $q_\perp$ behavior is present
in their sum $Q$ explicitly. The reason is that in this case all
individual large
(compared to $q_\perp$) contributions are mutually cancelled. The first step
is to use the Dirac equations $\hat p_1 u(p_1) = m u_1$,
$\bar u(p_1') \hat p_1' = m \bar u(p_1')$ and to rearrange
the amplitudes ${\cal{M}}_j$ of Eq.~(\ref{amplitudes}),
\begin{eqnarray} \label{eq:14}
  {\cal{M}}_1&=& \bar u(p_1') \Biggl\{
  \frac{s\beta_1'}{d_1}\hat e_2^*(\hat p_1-\hat k_1+m) \hat e_1^* \nonumber \\
  &-&  \frac{1}{d_1}\hat p_2\hat q \hat e_2^*
  (\hat p_1-\hat k_1+m) \hat e_1^* \Biggr\} u(p_1)\,,
\nonumber
\\
  {\cal{M}}_2&=& \bar u(p_1') \Biggl\{
  + \frac{s(1-x_1)}{d_1 d_2'} \hat e_2^*
  (\hat p_1-\hat k_1+m) \hat e_1^*
  - \frac{1}{d_2'} \hat e_2^* \hat p_2 \hat e_1^*
\nonumber
\\
  && +\frac{1}{d_1 d_2'} \hat e_2^* \hat q \hat p_2
  (\hat p_1- \hat k_1+m) \hat e_1^* \Biggr\} u(p_1) \,,
\\
  {\cal{M}}_3&=& \bar u(p_1') \Biggl\{
  \frac{s}{d'd_2'} \hat e_2^* (\hat p_1-\hat k_1+m) \hat e_1^*
  + \frac{s}{d'd'_2} \hat e_2^* \hat q \hat e_1^*
\nonumber
\\
  && + \frac{1}{d'd'_2} \hat e_2^*(\hat p_1'+\hat k_2+m)
  \hat e_1^* \hat q \hat p_2 \Biggr\} u(p_1) \,.
\nonumber
\end{eqnarray}
{}From these formulae it can be noted that the last terms in
${\cal{M}}_1, {\cal{M}}_2, {\cal{M}}_3$, up to terms of the order of
\begin{eqnarray*}
  \frac{m^2}{E^2} \,, \ \ \theta^2 \,, \ \
  \frac{m}{E} \,\theta\,,
\end{eqnarray*}
are proportional to $q_{\perp}$,
\begin{eqnarray} \label{eq:15}
  \hat{\tilde p}_2\hat q = \hat{\tilde p}_2 (\alpha_q \hat{\tilde p}_2 +
  \beta_q \hat p + \hat q_\perp) =
  \hat{\tilde p}_2\hat q_\perp= - \hat q \hat{\tilde p}_2 \,.
\end{eqnarray}
Next, one can see that the sum of the first three terms
in Eqs.~(\ref{eq:14}) is also proportional to $q_\perp$ since
(for more details see~\cite{DB})
\begin{eqnarray} \label{eq:16a}
  A \equiv \frac{b}{d_1} + \frac{1-x_1}{d_1 d_2'} + \frac{1}{d'd_2'} \,,
  \qquad A |_{q_{\perp}\to 0}=0\,.
\end{eqnarray}
Finally we consider the sum of the second terms of the quantities
${\cal{M}}_2, {\cal{M}}_3$ given in Eqs.~(\ref{eq:14}). Using the
relations~(\cite{DB}, Eq.(21)) and
\begin{eqnarray*}
(p_1' + k_1 + k_2)^2 = (p_1-k)^2 = m^2-{\mathbf {k}}^2-s\alpha_k \,,
\end{eqnarray*}
one immediately gets
\begin{eqnarray} \label{eq:17}
  -\frac{\hat{\tilde p}_2}{d_2'} + \frac{s(\alpha_q \hat{\tilde p}_2
  + \hat q_{\perp})}{d'd'_2} = \frac{s\hat q_{\perp}}{d'd'_2}
  + \frac{\hat{\tilde p}_2 {\mathbf {q}}^2}{d'd'_2} \,.
\end{eqnarray}
Therefore, from Eqs.~(\ref{eq:15}),~(\ref{eq:16a}),~(\ref{eq:17})
it is clearly seen that the property illustrated by Eq.~(\ref{eq:13})
\begin{eqnarray*}
  \left({\cal{M}}_1+{\cal{M}}_2+{\cal{M}}_3 \right)
  \vert_{q_{\perp}\to 0}=0
\end{eqnarray*}
is evidently satisfied and consequently the quantity
$Q=\sum_{j=1}^3 {\cal{M}}_j$ became a sum of terms explicitly
proportional to $q_{\perp}$,
\begin{eqnarray} \label{eq:18}
  Q&=&\bar u(p_1')\biggl\{A s \,\hat e_2^*
  (\hat p_1-\hat k_1+m) \hat e_1^*  \nonumber \\
  &-& \frac{1}{d_1} \hat{\tilde p}_2 \hat q_{\perp} \hat e_2^*
  (\hat p_1-\hat k_1+m) \hat e_1^*   \nonumber \\
  &-& \frac{{\mathbf {q}}^2} {d'd'_2} \hat e_2^* \hat e_1^* \hat{\tilde p}_2
  + \frac{s}{d'd'_2 } \hat e_2^* \hat q_{\perp} \hat e_1^*
\\ \nonumber
  &+& \frac{1}{d_1 d_2'} \hat e_2^* \hat q_{\perp} \hat{\tilde p}_2
  (\hat p_1-\hat k_1+m) \hat e_1^*  \\ \nonumber
  &+& \frac{1}{d'd'_2 }
  \hat e_2^* ( \hat p_1'+\hat k_2+m) \hat e_1^* \hat q_{\perp}
  \hat{\tilde p}_2\biggr\}u(p_1)\,.
\end{eqnarray}

Calculating the contribution of the trace ${\mathrm{Sp}}
\{p_1'Qp_1\tilde Q\}$
we neglect masses whose contribution to the quantity
$\Phi^{\gamma\gamma}$ may be restored using the general
prescription~\cite{ber}. The corresponding correction has the form:
\begin{eqnarray}
\Delta_m\Phi^{\gamma\gamma} &=& (1+{\cal P}_{12})\biggl\{ -\frac{4m^2}{d_1^{'2}}
\frac{x_2^2y_1(1+y_1^2)}{(1-x_2)^2} \\ \nonumber
&\times&\frac{{\mathbf{q}}^2}{({\mathbf{q}} - y_1{\mathbf{p}}_1')^2
({\mathbf{q}} - {\mathbf{p}}_1'/b)^2} \nonumber \\
&-& \frac{4m^2}{d_1^2}\frac{\beta_2^2z_1
(1+z_1^2){\mathbf{q}}^2}{({\mathbf{q}} - {\mathbf{p}}_1')^2({\mathbf{p}}_1' -
(1-\beta_2){\mathbf{q}})^2} \biggr\},
\end{eqnarray}
where
\begin{eqnarray}
y_1 = \frac{1 - x_2}{b},\quad \beta_2 = \frac{x_2}{1-x_1},\quad
z_1 = \frac{b}{1-x_1}.
\end{eqnarray}

\section*{Appendix C. Evaluation of 2-dimensional integrals }
\label{sec:AppC}
\setcounter{equation}{0}
\renewcommand{\theequation}{C.\arabic{equation}}

The azimuthal integration may be performed making use of the following equality:
\begin{eqnarray}
J_{12\ldots n}&=&\frac{1}{2\pi}\int_0^{2\pi}d\phi \prod\limits_i
[a_i+b_i\cos(\phi-\phi_i)]^{-1}
\nonumber \\
&=&\sum_{k=1}^n\frac{1}{r_k}\prod\limits_{j\not= k}^n\frac{b_k}{b_{kj}
+ \ii r_k\sin(\phi_k-\phi_j)}\,,
\end{eqnarray}
with
\begin{eqnarray*}
r_i&=&\sqrt{a_i^2-b_i^2},\qquad |a_i|>|b_i|, \\
b_{ij}&=&b_ia_j-b_ja_i\cos(\phi_i-\phi_j).
\end{eqnarray*}
It is curious to note that the absence of the imaginary part
provides an interesting algebraic identity.
For $n=2, n=3$ it looks
\begin{eqnarray}
J_{12}&=&\frac{1}{d_{12}}(\frac{b_1}{r_1}b_{12}+\frac{b_2}{r_2}b_{21}), \quad
d_{12}=a_{12}^2-r_1^2r_2^2,  \\ \nonumber
a_{12}&=&a_1a_2-b_1b_2\cos(\phi_1-\phi_2),\\ \nonumber
J_{123}&=&\frac{b_1^2}{r_1}\frac{a_{12}a_{13}-r_1^2a_{23}}{d_{12}d_{13}}
 + \frac{b_2^2}{r_2}\frac{a_{21}a_{23}-r_2^2a_{13}}{d_{12}d_{23}}  \\ \nonumber
 &+&\frac{b_3^2}{r_3}\frac{a_{31}a_{32}-r_3^2a_{12}}{d_{31}d_{32}}.
\end{eqnarray}
This form is convenient for a subsequent
integration over $\dd{\mathbf{k}}_1^2$.

\section*{Appendix D. NLO contributions from virtual and soft
photon emission }
\label{sec:AppD}
\setcounter{equation}{0}
\renewcommand{\theequation}{D.\arabic{equation}}

To avoid the misprints we use here the notations of the paper~\cite{heavy}
\begin{eqnarray}
s&=&d_1',\quad t=-d_1,\quad u=-Q^2, \\ \nonumber
s+t+u&=&q^2, \quad \tilde{f}(s,t)=f(t,s),\quad a=s+t, \\ \nonumber
b&=&s+u,\qquad c=u+t.
\end{eqnarray}
The quantities $\tau_{ij}$ encountered in the text (see Eq.~(\ref{loops}))
may be written as
\begin{eqnarray}
\tau_{11}&=& - G\left(1 + \frac{u^2}{s^2}\right)
- \tilde{G}\left(2+\frac{b^2}{t^2}\right)
+ 2 \biggl[\frac{b^2}{st} + \frac{2u}{a}  \nonumber \\
&+& \frac{2}{a^2}(u^2-bt)\biggr]l_{qu}
 + \frac{b^2}{tc^2}(2c+t)l_{qs} + \frac{2u-s}{s}l_{qt} \\ \nonumber
 &+& \frac{1}{q^2}\biggl[\frac{4}{a}(bt-u^2) - 4u - 2q^2 + t
 - \frac{b^2}{c}\biggr]\,, \\ \nonumber
\tau_{12}&=&\frac{c}{s^2}(u-s)G + \frac{1}{t^2}(uq^2-st)\tilde{G}
- 2\biggl[\frac{uq^2}{st} + \frac{2u-s+t}{a} \\ \nonumber
&+& \frac{2}{a^2}(u^2-cs)\biggr]l_{qu}
+ \frac{2c+t}{c^2}\left(s-\frac{u}{t}q^2\right)l_{qs} \\ \nonumber
&-& \frac{c}{bs}(2u-s)l_{qt}
+ \frac{1}{q^2}\biggl[\frac{4}{a}(u^2-cs) + 8u + 3t \\ \nonumber
&-& s + \frac{2}{c}us\biggr],
\end{eqnarray}
and the additional notations look
\begin{eqnarray}
l_{qu}&=&\ln\frac{q^2}{u},\quad l_{qs}=\ln\frac{-q^2}{s},\quad
l_{qt}=\ln\frac{q^2}{t},\quad l_{ut}=\ln\frac{u}{t}\,, \nonumber \\
G&=&l_{qu}(l_{qt} + l_{ut}) + 2\Li\left(1-\frac{t}{q^2}\right)	 \\ \nonumber
&& - 2\Li\left(1-\frac{q^2}{u}\right)
- 2\Li(1).
\end{eqnarray}

\section*{Appendix E. Semi-collinear kinematics of pair creation }
\label{sec:AppE}
\setcounter{equation}{0}
\renewcommand{\theequation}{E.\arabic{equation}}

The matrix element in the kinematics~(\ref{eq:1}) may be put in a form
(we extract the coupling constant):
\begin{equation}
M^{(1)}=\frac{1}{q_1^2}J_\nu I_\mu g^{\mu\nu},\quad
J_\nu=\bar{u}(p_-)\gamma_\nu u(p_1),
\end{equation}
where the current $I$ describes a pair production by the photon
with momentum $q_1$ off a proton.
Using the Sudakov form of the 4-vectors $p_-$ and $q$ with basic 4-vectors
$p_1$ and $p_2$,
$$
p_-=\alpha_-\tilde{p}_2+\beta_-\tilde{p}_1+p_{-\bot},\quad
q =\alpha_q\tilde{p}_2+\beta_q\tilde{p}_1+q_\bot,
 $$
the representation of the metric tensor
$$
g_{\nu\mu}=g_{\nu\mu\bot}+\frac{2}{s}p_{2\nu}p_{1\mu}
$$
and the gauge condition
$$
Iq=I(\beta_qp_1+q_\bot)=0,\quad \beta_q+\beta_-=1,
$$
we obtain
for the matrix element squared and summed over spin states of electron:
\begin{equation}
\sum|M^{(1)}|^2=\frac{1}{(q_1^2)^2}\Biggl[-2q_1^2{\mathbf{I}}^2
+ \frac{8}{\beta_q^2}\biggl({\mathbf p}_- {\mathbf{I}}\biggr)^2\Biggr].
\end{equation}
To calculate the quantity ${\mathbf{I}}^2$, we again present it in a form
\begin{eqnarray}
I&=&e_{q_1}{\mathbf{I}} = e_{q_1}^\mu e_q^\nu \frac{2s|\vec{q}|}
{q^2s_1}p_{2\rho}Y_\rho \bar{u}(p_1')O_{\mu\nu}v(p_+), \nonumber \\
s_1&=&(p_2+q_1)^2,\qquad Y_\rho=\bar{u}(p_2)\Gamma_\rho u(p_2').
\end{eqnarray}
The phase volume is transformed the way to take the following form
\begin{equation}
\dd\Gamma_4=(2\pi)^{-8}\frac{1}{8s\beta_-\beta_+ b }\dd^2q\dd^2p_-\dd\beta_-.
\end{equation}
Using $$\sum|\bar{u}(p_1')O_{\mu\nu}v(p_+)e_{q_1}^\mu e_q^\nu|^2=
8\biggl[\frac{b}{\beta_+} + \frac{\beta_+}{b}\biggr],$$
we obtain the result for the cross section given in the text.

For the kinematics of bremsstrahlung mechanism the matrix element
has a form
\begin{equation}
M^{(2)}=\frac{1}{k_1^2}I_\mu J_\nu g^{\mu\nu},\quad k_1=p_+ + p_1'.
\end{equation}
Here it is suitable to
use alternative basis vectors of Sudakov parameterization
\begin{eqnarray}
p_+&=&\alpha_+ q+b_+\tilde{p}_1'+p_{+\bot}, \
k_1=a_1 q+b_1\tilde{p}_1'+k_{1\bot}, \\ \nonumber
g_{\mu\nu}&=&g_{\mu\nu\bot} + \frac{2}{\tilde{s}}q^\nu p_1^{'\mu}, \quad
k_1^2=\frac{{\mathbf{p}}_+^2 + m^2 b_1^2}{b_1 - 1}>0.
\end{eqnarray}
Quite the same manipulations give
$$
\sum|M^{(2)}|^2=2k_1^2{\mathbf{I}}^2 - \frac{8}{b_1^2}\biggl
({\mathbf{k}}_1{\mathbf{I}}\biggr)^2.
$$
Performing the integration over $\dd^2(p_+)_\bot$ to
a logarithmic accuracy
and expressing the parameter $b_1$ in terms of the standard Sudakov
decomposition with basic 4-vectors $p_1,p_2$
$$
b_1=\frac{1-\beta_-}{b}\,,
$$
we immediately obtain the result given in the text.


\end{document}